\documentclass[twocolumn,showpacs,preprintnumbers,amsmath,amssymb, epsfig]{revtex4}

\usepackage{graphicx}
\usepackage{dcolumn}
\usepackage{bm}
\usepackage{epstopdf}

\begin{document}
\preprint{}

\title{Massless spacetime: On spacetime geometry above the electroweak symmetry breaking scale}

\author{Ka\'{c}a Bradonji\'{c}}
\affiliation{%
Department of Physics, Boston University, Boston, MA 02215 USA
}%

%

\date{May 27, 2009}

\begin{abstract}
One of the deepest insights from the general theory of relativity is the relational nature of spacetime. While it is a generally agreed on that the nature of spacetime must be drastically different at the Planck scale, it has been a common practice to assume a full pseudo-Riemannian geometry of spacetime no matter what the physical fields present are, and to apply such geometry to cosmology. In the spirit of preserving the relational description of spacetime, we initiate a discussion on the applicability of general relativity all the way to Planck scale, and propose the possibility that the full pseudo-Riemannian geometry emerges as late as the time of electroweak phase transition when spacetime acquires the projective structure necessary to describe the motions of massive particles. A brief discussion of possible properties of spacetime prior to electroweak symmetry breaking is followed by an exploration of plausible experimental indications that such transition is real. 
\end{abstract}

\maketitle

\section{Introduction}
The general theory of relativity (GR) radically changed the way we understand space and time. In addition to unifying space and time, Einstein realized that one of the biggest implications of his theory is that space and time are real only insofar as they are relations between physical entities (particles or fields.) He noted that as a consequence of the general covariance of GR, spacetime measurements are simply measurements of the \emph{coincidences} of particles, be they material points of the measuring rods, clock hand and the points of the clock dial, or observed point events happening at the same place and the same time \cite{Einstein1952}. Much has been written about the intimate connection between the nature of spacetime and the physical fields, as well as reality of some spacetime structures, and these topics continue to be studied \cite{Stachel}.  

Spacetime in GR is described by a four-dimensional differentiable pseudo-Riemannian manifold $M$ equipped with a metric tensor $g_{\mu \nu}$ and a metric-compatible connection $\{^{\sigma} _{\mu \nu}\}$, also known as Christoffel symbols, which are functions of the metric and its derivatives  \cite{Wald}. It is widely accepted that the nature of spacetime at very high energies ought to be different.  Numerous approaches to the problem of quantum gravity, such as loop quantum gravity  \cite{Thiemann}, spin-foams \cite{Rovelli}, causal set theory \cite{Sorkin}, causal dynamical triangulations \cite{Loll}, quantum graphity \cite{Fotini}, just to name a few, start off with a pre-spacetime structure in the hope of recovering the full pseudo-Riemannian manifold in some limit. However, the assumption that spacetime can be described by a pseudo-Riemannian manifold (flat or curved) all the way up to Planck scale has been common practice in the applications of GR to cosmology and particle physics, including the Big Bang theory \cite{Wald}, inflation \cite{Guth}, as well as  the Standard Model (SM) \cite{GWS}.

Taking to heart Einstein's deep insight that the nature of spacetime is genuinely relational, it seems quite reasonable to ask whether there are geometrical structures of a pseudo-Riemannian manifold which are extraneous if we consider the existence of only certain kinds of physical fields. If we assume that spacetime is fully described by a pseudo-Riemannian manifold no matter what physical fields exist in it, then we are allowing it to have degrees of freedom and symmetries that may not, even in principle, be measurable. If, on the other hand, we accept that properties of spacetime are meaningful only as a description of relations among physical entities and ought to be \emph{in principle} measurable, then we have to ask which structures of spacetime are justifiably presumed to exist when only specific kinds of fields are present. We emphasize that the requirement that a spacetime structure be ``in principle measurable" is simply a requirement that the spacetime structure in question \emph{could} manifest itself through behavior of existing physical fields, and not that it \emph{does} \cite{Malament}.

In this paper we explore the issue at hand by considering the description of spacetime with an energy content higher than that at which electroweak symmetry breaking (EWSB) occurs. A pseudo-Riemannian manifold seems appropriate for modeling spacetime which accommodates the massless and massive physical fields of the SM. But according to the SM, prior to EWSB there were no massive particles - all particles acquired mass via the Higgs mechanism at the EWSB energy. Assuming that there were no massive particles prior to the electroweak phase transition, and taking to heart the relational nature of spacetime, the question arises whether it is meaningful to endow spacetime at energies higher than the EWSB energy with the full pseudo-Riemannian geometry. More specifically, in a world in which all particles are massless and time-like curves physically never manifest themselves through motion of massive particles, we ask if it is meaningful to assume that such curves exist. A more formal and illuminating form of this question can be posed in the language of conformal and projective structures which we describe in the following section. The purpose of this paper is not to present a fully worked out description of spacetime at high energies, but only to initiate a discussion on the subject  and point out some interesting questions that deserve consideration. 

\section{Conformal and projective structures}
In the Palatini formalism of GR, the metric $g_{\mu\nu}$ and the affine connection $\Gamma^{\sigma}_{\mu\nu}$ are treated as independent variables and their compatibility is derived as a result of the vanishing variation of the gravitational vacuum action with respect to the connection. The metric tensor can be further divided into a conformal structure and the four-volume element, and the affine connection can be broken up into the projective structure and a choice of a curve parametrization \cite{Schouten}.

A conformal structure $\mathcal{C}$ on a manifold is mathematically described by an equivalence class  $\{\Omega^{2} g_{\mu\nu}\}$  of metrics $g_{\mu\nu}$ related by a positive factor $\Omega^{2}$, defined at every point of the manifold. Alternatively, it can be described by a tensor density field 
\begin{equation}
\tilde{g}_{\mu\nu}=(-g)^{-\frac{1}{4}}g_{\mu\nu},
\end{equation}
where $g$ is the determinant of the metric. $\mathcal{C}$ is invariant under the conformal transformations of the metric
\begin{equation}
g_{\mu \nu}\longrightarrow \Omega^{2} g_{\mu\nu},
\end{equation}
where $\Omega^{2}$ is a positive factor. 

A projective structure $\mathcal{P}$ is mathematically represented by projective parameters
\begin{equation}
\Pi^{\sigma}_{\rho \lambda}=\Gamma^{\sigma}_{\rho \lambda}-\frac{1}{5}(\delta^{\sigma}_{\rho}\Gamma_{ \lambda}+\delta^{\sigma}_{\lambda}\Gamma_{ \rho}),
\end{equation}
where $\Gamma_{\rho}=\Gamma^{\sigma}_{\rho \sigma}$ is the trace of the affine connection. $\mathcal{P}$ is invariant under the projective transformations of the affine connection
\begin{equation}
\Gamma^{\sigma}_{\rho \lambda} \longrightarrow \Gamma^{\sigma}_{\rho \lambda}+\frac{1}{5}(\delta^{\sigma}_{\rho}p_{\lambda}+\delta^{\sigma}_{\lambda}p_{\rho}),
\end{equation}
where $p_{\rho}$ is an arbitrary one-form. 

Both $\mathcal{C}$ and $\mathcal{P}$ have clear physical interpretations and can be constructed at each point from curves traversable by light rays (or free massless particles) and free massive particles respectively \cite{Ehlers1972, Pirani1973}. Here ``free"  means not under influence of anything but gravitational effects. These particles are considered to be test particles, and no consideration of their spin or other quantum numbers is made. The full pseudo-Riemannian geometry of spacetime can be recovered by introducing $\mathcal{C}$ and $\mathcal{P}$ on a differentiable manifold and imposing two compatibility conditions between them. All the considerations of this formalism are local and assume that the manifold and the curves in question are differentiable. Here we just summarize in broad brushstrokes the key steps of such axiomatic construction:
\begin{itemize}
\item {The propagation of light determines at each point of spactime an infinitesimal null cone and hence defines $\mathcal{C}$ on $M$. This structure allows for distinction among time-like, null, and space-like vectors, directions, curves, etc. at a point. Light rays are represented by null geodesics which are null curves contained in null hypersurfaces.}
\item{The motions of freely falling massive particles determine a family of preferred unparametrized time-like curves at each point, and such a family at each point of $M$ defines $\mathcal{P}$ on $M$. World lines of freely falling particles are said to be $\mathcal{C}$-time-like geodesics of $\mathcal{P}$.}
\item{Requiring two compatibility conditions: 1) that the null geodesics are also geodesics of $\mathcal{P}$, and 2) that the ticking rate of a clock is independent of its history leads to the full pseudo-Riemannian geometry.}
\end{itemize}
The axiomatic construction of pseudo-Riemannian geometry from $\mathcal{C}$ and $\mathcal{P}$ and the two compatibility conditions illuminates the correspondence between the geometrical structures of spacetime and the physical fields through whose behavior these structures manifest themselves. Aware of the physical motivation of $\mathcal{C}$ and  $\mathcal{P}$, we are now obliged to inquire whether it is justified to assume that these spacetime structures are always present no matter what physical fields exist. We initiate the discussion by considering the existence of the projective structure in a spacetime above the EWSB scale.

\section{Spacetime geometry above EWSB scale}
Starting point of this discussion is the assumption that the spacetime is truly relational, and that any of its presumed geometrical structures ought to have, at least \emph{in principle} possible physical manifestation \cite{Smolin2005}. It is commonly accepted that spacetime is of fundamentally different nature at Planck scale, as at those energies significance of gravitational quantum effects becomes large and GR is not the correct description of the world. Meaning of time without a mass to fix a length scale \cite{Penrose2008}, as well as the significance of time in quantum gravity \cite{Isham}, have been analyzed. It has even been suggested that time may not be meaningful as early as the quark-gluon phase transition \cite{Rugh}. The applicability of GR to cosmology and definition of cosmological time has also been under discussion \cite{Barbour}. We assume that GR can be applied to cosmology, and consider a somewhat different question. 

According to the SM, our universe underwent several phase transitions, one of which is the electroweak symmetry breaking \cite{Boyanovsky, Dawson}. Prior to the EWSB, the Higgs field had a vanishing vacuum expectation value (vev) and all the particles were massless. Given that all the fields of the SM were massless prior to electroweak symmetry breaking and that the projective structure of spacetime is motivated by the existence of massive particles, we are obliged to ask:

\emph{Is it meaningful to maintain that spacetime has projective structure at energies above the electroweak symmetry breaking scale?} 

Assuming that the time-like curves of $\mathcal{P}$ manifest themselves exclusively by motions of massive particles, and that spacetime structures should have observable manifestations, maintaining the existence of $\mathcal{P}$ without massive particles to indicate its presence is not justified. Looking at this another way, if we were to attempt the axiomatic construction, analogous to the one described in the previous section, of the classical spacetime with an energy content higher than the EWSB energy, we would not be able to recover a full pseudo-Riemannian geometry.  Current cosmological theories assume the presence of the full pseudo-Riemannian geometry all the way up to Planck scale - in the light of our discussion, this assumption should be subject to more scrutiny.

Instead of just questioning the meaning of length and time in this ``massless spacetime", one could take a step further and suggest that $\mathcal{P}$ is superfluous prior to the EWSB.  What is suggested here is not merely that any theory describing the world at such energies must be conformally invariant, and hence scale free, but that the spacetime \emph{itself} is devoid of projective structure. Since $\mathcal{P}$ is the key component of pseudo-Riemannian geometry, this would mean that prior to the EWSB, spacetime had a different geometry. As electroweak symmetry phase transition gave particles their masses, massive particles gave rise to the projective structure breaking the conformal symmetry of spacetime and resulting in the emergence of a full pseudo-Riemannian geometry. This description could apply to the universe as a whole, or to a small region of spacetime with an energy content high enough to recreate the conditions prior to EWSB.  Only way out of having to drop $\mathcal{P}$ above the EWSB energy is to find some manifestation of time-like curves that does not involve the motion of massive particles, and the present author is not familiar with any such thing. 

It has been suggested that at early times and high temperatures the masses of all known particles could be considered ``effectively zero'' and that one might imagine allowing the metric to be conformally rescaled by a very small factor, making the overall metric tend to zero \cite{Penrose2006}. In this case, the test particles never go to exactly zero mass and at all stages the full conformal and projective structures can be meaningfully defined - something not the case if one enters a phase where there are {\it no} massive particles at all that can be used to construct the full metric geometry.

It is interesting to consider what spacetime would look like without $\mathcal{P}$.
Without a mass to set the length scale, there is no meaningful way to define a clock, so the notion of time is as fuzzy as is the notion of length. Since massless particles follow null curves (geodesics or not), only null curves are observable and all physical events are connected by such null curves. There is no question of \emph{how} two events are separated (time-like, null, or space-like), but simply \emph{if} they are connected by a null curve or not.  In essence, the pseudo-Riemannian geometry of spacetime reduces to a conformally invariant geometry. 

\begin{figure}
\includegraphics{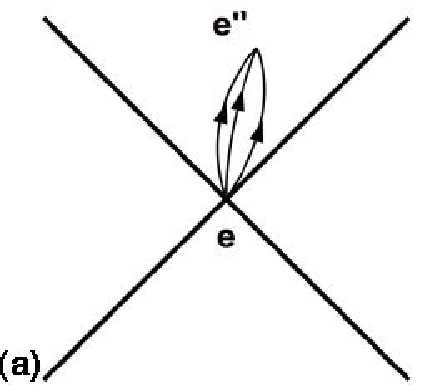}\includegraphics{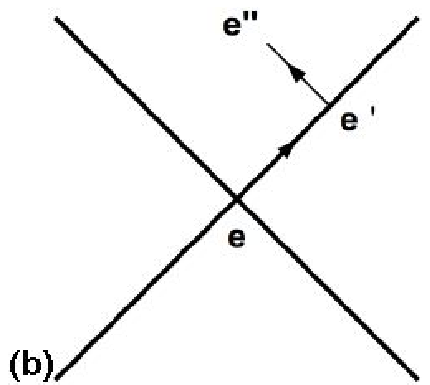}
\caption{\label{fig:fig1}Events $e$ and $e^{''}$ connected: (a) with the projective structure, and (b) without the projective structure at $e$ and $e^{''}$.}
\end{figure}
Figure \ref{fig:fig1}(a) shows two time-like separated events $e$ and $e^{''}$ of the pseudo-Riemannian spacetime connected by time-like curves. Taking away $\mathcal{P}$ amounts to forbidding such a communication along time-like curves and leaves $e$ and $e^{''}$ in Fig.\ref{fig:fig1}(b) with a ``second degree" of connection via a bouncing massless particle, resembling the Feynman checkerboard idea \cite{Feynman}. The null curve connecting $e$ and $e''$ is discontinuous and consists of two null segments.  Since there is no time-like curve between $e$ and $e''$, it is not clear if one can define the separation of these two events to be anything else but null. If this is indeed the case, all events of the massless spacetime have a separation of $\int ds=0$. Massless spacetime still maintains a part of causal structure, which suggests interesting connections to the theory of causal sets. 

Although taking away $\mathcal{P}$ from the pseudo-Riemannian spacetime leaves us a spacetime endowed with $\mathcal{C}$, and such spacetime is a mathematical possibility, it may not be physically meaningful. Even though conformal curvature, and ratios of lengths and angles are well defined in a conformally invariant spacetime \cite{Pirani1966}, it is an open question how, if at all, these things could be measured within the framework of SM and prior to EWSB. Unless a way to conduct such measurements \emph{in principle} is found,  we may have to consider a more primitive model of spacetime above EWSB scale.

\section{Observable consequences}
One ought to consider whether or not a change in geometry of spacetime at the EWSB scale has testable consequences. Certainly without time, it is difficult to deal with dynamics. Nevertheless, there are situations in which one might expect such a transition to be observable.

Consider a region of spacetime $S$ whose temperature is slowly raised. At some point, the system will reach a phase where the Higgs field no longer has a nonvanishing vev. Let us consider this to be a sharp yes/no decision and the presence or absence of the Higgs vev to also correspond to whether the region $S$ admits $\mathcal{P}$ or not. Such regions might be expected in at least two physical situations. 

One instance is the early universe with size $R$ set to be the radius of the universe at the temperature  $T_{w}$ when the electroweak phase transition is supposed to have occurred. Within the framework of ideas proposed above, one might argue that $\int{ds}=0$ between all points would lead to a perfect thermal equilibrium at the moment of symmetry breaking (which indeed, would also be the emergence of time for the system's evolution). In the early universe scenario, one might expect a similar perfectly isotropic thermal distribution or scale invariant fluctuations paralleling the state of things looked for in inflationary cosmology which, however, assumes the full pseudo-Riemannian spacetime all the way down to the Planck scale and postulates an additional scalar field. Possible signatures might be found in the CMB spectrum \cite{Bradonjic}. To support the claim that spacetime itself is conformally invariant, and not just that theory describing the physical fields has conformal symmetry, one may have to look for a framework which makes the distinction clear. A possible way would be to understand how the emergence of the full pseudo-Riemannian geometry would affect gravitational entropy and how such transition would manifest itself in the state of the current universe \cite{Penrose2006}. 

Another place to look for a confirmation that this transition actually takes place could be in the case of heavy ion collisions or cosmic rays, where one might expect a very isotropic, thermally uniform distribution of outgoing particles. The distribution would be as uniform as allowed for by thermodynamic fluctuations in temperature  \cite{Stodolsky}, (remember that the Planck spectrum is essentially one that arises from conformal invariance). However, it is not clear how in this case the conformal symmetry of spacetime could be decoupled from the conformal symmetry of the physical theory describing the heavy ion collisions.

\section{Conclusions}
In the spirit of preserving the relational nature of spacetime, it may be important to revisit our assumptions about the validity and scope of general relativity when no massive fields are present. Having no physical motivation, the projective structure as described in \cite{Ehlers1972}, and even some aspects of a conformal structure seem superfluous in the description of the spacetime prior to the electroweak symmetry breaking. In a sense, what is proposed here is that there may be two scales in gravity. The usual Planck scale should be either supplanted or supplemented by an additional scale which is set by the electroweak symmetry breaking energy at which sufficient physical structure appeared to allow the full conformal and the projective structures to appear. If this is the case, there could be genuine new gravitational physics at the electroweak scale possibly accessible to particle physics or cosmic ray experiments and cosmological observations right now, and without the need to postulate large extra dimensions \cite{Arkani1999, Arkani2000,Randall,Shiu}. The considerations of this paper, however,  do not depend on the details of the process by which particles acquire mass, but assume only that there was a time when all particles were massless, and that such a process took place. Whatever energy this process may occur at is the energy at which we should question the nature of spacetime and the validity of general relativity. We note that this line of reasoning would not have occurred to Einstein since in his time, massive particles were considered to have mass at all energy scales - it is only the advent of the SM that makes this line of thought reasonable. The hope of the present author is that the thoughts and questions posed in this paper will initiate further discussions on the matter.

\begin{acknowledgments}I would like to thank John Stachel for introducing me to the work on conformal and projective structures, and discussion on relational nature of spacetime. I am also grateful to John Swain for reading an early draft of this paper, as well as for the discussion on possible physical consequences of scale dependent spacetime geometry. 
\end{acknowledgments}

 
\end{document}